\documentstyle[psfig]{caosp}
\begin{document}
\def\Hm{\langle H\rangle}
\def\Hs{(\Hsq+\Hzsq)}
\def\Ht{\Hs^{1/2}}
\def\Hsq{\langle H^2\rangle}
\def\Hzsq{\langle H_z^2\rangle}
\def\HgII{\hbox{{\rm Hg}~{$\scriptstyle {\rm II}$}}}
\def\PtII{\hbox{{\rm Pt}~{$\scriptstyle {\rm II}$}}}
\def\AuII{\hbox{{\rm Au}~{$\scriptstyle {\rm II}$}}}
%
\pubyear{1998}
\volume{27}
\firstpage{344}
\hauthor{S. Hubrig {\it et al.}}
\title{Isotopic composition of Hg and Pt in slowly rotating HgMn stars}
\author{S. Hubrig \inst{1} \and F. Castelli
\inst{2}   \and G. Mathys \inst{3}}
\institute{University of Potsdam, Am Neuen Palais 10, D-14469 Potsdam,  Germany
\and CNR-Gruppo Nazionale Astronomia and Osservatorio Astronomico, Via
G. Tiepolo 11, I-34131 Trieste, Italy  \and 
European Southern Observatory, Casilla 19001, Santiago 19, Chile}

\date{\today}
\maketitle
\begin{abstract}
Preliminary results from a study of the isotopic
compositions of the elements Hg and Pt in a number of HgMn stars
are presented. This work represents an improvement over previous studies 
thanks to the very high spectral resolution available (R = 118 000) and to 
the new information on wavelengths and atomic structure of \HgII\ and \PtII.
\end{abstract}
\section{Introduction}
\label{intr}
Almost all HgMn stars exhibit a strong absorption feature at
3984~\AA, which has been identified as a line of \HgII.
The wavelength at 
which this line is observed depends on the isotopic mix of Hg
(White et al. 1976), which 
ranges from the terrestrial mix to nearly pure $^{204}\!$Hg.
Mercury is not the only very heavy element observed in HgMn stars.
Lines of \PtII\ (Dworetsky \& Vaughan 1973) and \AuII\ (Guthrie 1985) are 
also observed.
Dworetsky \& Vaughan (1973) studied the \PtII\ $\lambda\,4046$ line in a 
sample of nine HgMn stars. This line is the strongest Pt line at optical
wavelengths, and in the nine stars studied it is shifted toward
longer wavelengths by 0.04 to 0.09~\AA, with respect to the centroid of 
the terrestrial platinum line. These shifts are interpreted as an
isotopic effect. The corresponding anomalies are
analogous to those found for Hg, in the sense that the
heavier isotopes tend to dominate in cooler stars.
Neither radiatively driven diffusion nor any other theory 
until now can account
satisfactorily for the variations in the Hg and Pt isotope mix 
among the HgMn stars (Leckrone et al. 1993).

The main purpose of the work reported here was to provide additional
observational constraints to guide the theorists in the understanding
of the isotopic anomalies in HgMn stars, improving upon previous studies
through the much better data quality obtainable now.
Thanks to the availability of new laboratory 
measurements of isotope shifts in \PtII\ (Engleman 1989) it became possible
to identify more definitely the \PtII\ isotopes.

\begin{table}[t]
\small
\begin{center}
\caption{Isotopic compositions}
\label{t1}
\begin{tabular}{lrrrrrr}
\hline\hline
Star & {Terrestrial} & {$\chi$~Lup} & {HR~7775} & {HR~1800} & 
{74 Aqr} & {HR 6520}\\
       &{abundance}\\
\hline\hline
$T_{\rm eff}$ (K)  &&10680             &10830         &11070     &11880  &13250\\
$\log g$     &&3.99              &4.11          &3.75       &4.03  &4.17\\
\hline\hline
{[Hg]}              && +5.45  & +5.65        & +5.25    &        & +5.10\\
\hline
&\multicolumn{6}{c}{Hg isotopic structure (\%)}\\
\hline
196 & 0.15 & 0.00 & 0.00 & 0.00 & 0.00 & 0.00 \\       
198 & 9.97 & 0.00 & 0.00 & 0.16 & 7.02 & 4.10 \\
199a & 7.14 & 0.00 & 0.00 & 0.32 & 6.19 & 10.99 \\
199b & 9.71 & 0.00 & 0.00 & 0.44 & 9.29 & 14.00 \\
200 & 23.09 & 0.00 & 0.00 & 2.97 & 26.79 & 28.97 \\
201a & 4.80 & 0.00 & 0.10 & 1.77 & 6.01 & 8.61 \\
201b & 8.30 & 0.00 & 0.20 & 3.06 & 9.10 & 8.93 \\
202 & 29.80 & 1.00 & 49.70 & 50.58 & 26.79 & 22.49 \\
204 & 6.86 & 99.00 & 49.70 & 40.74 & 8.81 & 1.80\\
\hline\hline
{[Pt]}                && +4.00           & +4.69         & +3.30\\  
\hline
&\multicolumn{4}{c}{Pt isotopic structure (\%)}\\
\hline
194 &32.90& 0.00& 0.00&  0.00 \\
195b&18.78& 0.00& 7.50&  0.00 \\
195a&13.15& 0.00&10.00&  0.00 \\
196 &25.20&10.00&55.00&  0.00 \\
198 & 7.19&90.00&27.50&100.00 \\
 \hline\hline
\end{tabular}
\end{center}
\end{table}

\section{Observations and spectrum synthesis}

Spectra were obtained with the ESO 1.4~m Coud\'e Auxiliary Telescope and
the Coud\'e Echelle Spectrograph Long Camera at a resolving power  
R = $\lambda/\Delta
\lambda=118\,000$ and ${\rm S/N}\geq250$. The observed wavelength ranges
are 3965--4000~\AA\ and 4018--4035~\AA.
Synthetic spectra and model atmospheres were computed with the SYNTHE 
and ATLAS9 codes (Kurucz 1997), respectively. 
A code similar to the 
TEFFLOGG code of Moon \& Dworetsky (1985), but based on new
computed uvbybeta indices, was used to obtain the stellar parameters
(Castelli \&\ Kurucz 1994). 
Observed indices were taken from the Mermilliod, Mermilliod,
\& Hauck  catalogue (1997) and were dereddened using the 
UVBYBETA code of Moon (1985).
For all the stars we assumed zero microturbulent velocity, while the
rotational velocity was derived from the comparison of the observed and
computed spectra, after having degraded the computed spectra for the
broadening due to the instrumental profile.
For the whole transition of \HgII\ 3983~\AA\  we adopted $\log gf=-1.73$ 
(Dworetsky 1980).  For each isotopic and hyperfine component this
value was scaled in agreement with each observed relative intensity.
For reference, the terrestrial intensities from Kurucz (1993) and 
from Smith (1997) were adopted.
For the transitions of \PtII\ 4023.8, 4034.2, and 4046.4~\AA\ we
adopted log $gf=-2.61, -2.09$, and $-1.19$, respectively 
(Dworetsky \& Vaughan 1973). The isotopic and hyperfine
shifts and intensities were either taken directly
from Engleman (1989) or were derived from Engleman (1989)
and Kalus et al. (1997). 
One of the programme stars with very sharp lines is the double-lined 
spectroscopic binary $\chi$~Lup. An updated version
of the BINARY code of Kurucz (1993, CD-ROM 18) has yielded as final 
computed spectrum the spectrum resulting from the contribution of both
components.
The atmospheric parameters of the secondary star and ratio of
the radii of the primary to the secondary stars in $\chi$~Lup were taken
from Wahlgren et al. (1994).

For five programme stars with extremely sharp spectral lines 
($v\,\sin i < 3$~km/s) 
the isotopic composition could be studied in greater detail.
The results are summarized in Table~1. 
All stars have Hg overabundance by more than 5~dex compared with the solar
abundance. The largest overabundance of Pt (4.69~dex) was found in
the star HR~7775. 
No star shows terrestrial isotopic proportions. The most pronounced 
deviation from the terrestrial composition is found in the stars $\chi$~Lup
and HR~7775, which are the coolest ones in our sample. 

The large overabundances of Hg and Pt and the star-to-star variations in their
isotopic composition clearly pose a challenge to
any theory aimed at explaining the origin of chemical peculiarities.


\begin{thebibliography}{}
\bibitem{}Castelli F., Kurucz R.L.: 1994, Poster paper presented
  at the JD No.~12 of the 22nd IAU General Assembly
\article{Dworetsky M.M.}{1980}{\aaa}{84}{350}
\article{Dworetsky M.M., Vaughan A.H.}{1973}{\apj}{181}{811}
\article{Engleman R.J.}{1989}{\apj}{340}{1140}
\article{Guthrie B.N.G.}{1985}{\mnras}{216}{1}
\bibitem{} Kalus G., Johansson S., Wahlgren G.M., Leckrone D.S., Thorne A.P.,
Brandt J.C. {\it Astrophys. J.}, in press 
\bibitem{}Kurucz R.L.: 1997, private communication
\article{Kurucz R.L.}{1993}{Phys. Scripta}{T47}{110}
\inproceedings{Leckrone D.S., Wahlgren G.M., Johansson S., Adelman S.J.}
{1993}{Peculiar Versus Normal Phenomena in 
A-Type and Related Stars}{M.M. Dworetsky, F. Castelli \& R. Faraggiana}{ASP 
Conf. Series}{44}{42}
\article{ Mermilliod J.C., Mermilliod, M., Hauck, B.}{1997}{\aaas}{124}{349}
\bibitem{}Moon T.T.: 1985, {\it Comm. Univ. London Obs.}{\bf 78}
\article{Moon T.T., Dworetsky M.M}{1985}{\mnras}{217}{305}
\article{Smith K.C.}{1997}{\aaa}{319}{928}
\article{Wahlgren G.M.,  Adelman S.J., Robinson R.D.}{1994}{\apj}{434}{349}
\article{White R.E., Vaughan A.H., Preston G.W., Swings J.-P.}{1976}{\apj}
{204}{131}
\end{thebibliography}
\end{document}